\documentstyle{article}

\newcommand{\be}{\begin{equation}}
\newcommand{\ee}{\end{equation}}

\title{\bf  Stability of the Mezard-Parisi solution for random manifolds}

\author{
 \\  
  {D. M. Carlucci}               \\
  {\small\it Scuola Normale Superiore di Pisa}  \\[-0.2cm]
  {\small\it Piazza dei Cavalieri}        \\[-0.2cm]
  {\small\it Pisa 56126, Italy}          \\
  {\small Internet: {\tt CARLUCCI@UX2SNS.SNS.IT}}     \\ 
 \\      
  { C.De Dominicis}              \\
  {\small\it S.Ph.T., CE Saclay } \\[-0.2cm]
  {\small\it 1991 Gif sur Yvette , France}          \\
  {\small Internet: {\tt CIRANO@AMOCO.SACLAY.CEA.FR}}     \\
 \\  
  { T. Temesvari}              \\
  {\small\it Institute for Theoretical Physics} \\[-0.2cm]
  {\small\it E\"otv\"os University } \\[-0.2cm]
  {\small\it 1088 Budapest, Hungary}          \\
  {\small Internet: {\tt TEMTAM@HAL9000.ELTE.HU}}     \\
       }

\begin{document}

\maketitle

\thispagestyle{empty}

\abstract{The eigenvalues of the Hessian associated with random manifolds 
	  are constructed for the general case of $R$ steps of replica 
	  symmetry breaking. For the Parisi limit $R\to\infty$ (continuum 
	  replica symmetry breaking) which is relevant for the manifold 
	  dimension $D<2$, they are shown to be non negative.}

\clearpage

\thispagestyle{empty}

         \centerline{\bf Resum\'e}

	  {Les valeurs propres de la Hessienne, associ\'ee avec une 
	  variet\'e aleatoire, sont construites dans le cas general de 
	  $R$ \'etapes de brisure de la symmetrie des repliques. Dans la 
	  limite de Parisi, $R\to \infty$ (brisure continue de la 
	  symmetrie des repliques) qui est pertinente pour la 
	  dimension de la variet\'e $D<2$, on montre qu'elles sont 
	  non negative.}

\vspace{5cm}


\clearpage

\section{Introduction}

The behaviour of fluctuating manifolds pinned by quenched random 
impurities encompasses a rich variety of physical situations, 
from the directed polymers in a random potential (with a manifold 
dimension $D=1$) to interfacial situations ($D=d-1$, $d$ being the 
total dimension of the space). The directed polymer problem, 
for example, is itself related to surface growth, turbulence 
of the Burgers equation and spin glass. The literature concerning 
various aspects is wide and florishing, a good deal of it being 
quoted in some recent reviews\cite{Natterman, Halpin}. 
The approaches taken have 
been quite diverse and it is not the purpose of this paper to describe 
them. This paper is a first step in an approach to random manifolds 
that uses field theoretic techniques in the replica formalism. 
Field theoretic techniques have been used with success (in their 
functional renormalization group variety) by Balents and 
Fisher\cite{Balents_1}
to describe the large $N$ ($N=d-D$ is the transverse dimension) for the 
$4-\epsilon$  dimensional manifold. In the context of directed polymers 
Hwa and Fisher\cite{Hwa}, using more conventional field theory, have 
emphasized the role of large but rare fluctuations. Alternately 
a unified description of the general random manifold has been 
developed by M\'ezard and Parisi\cite{Mezard_1}(MP), where, {\it via} the 
use of the replica technique, they have laid down the first step 
(mean field and self-consistent Hartree-Fock level) of a systematic 
field theory $1/N$ expansion\footnote{The intricacies of the first 
loop correction were briefly considered in further work\cite{Mezard_2}}. 
This is the approach we want to pursue here 
and this paper may be considered as a first sequel to MP. 

In this paper we study the eigenvalues of the Hessian associated with the 
$1/N$ expansion  as laid down by MP. We do it in a general form by 
keeping discrete the $R$ steps of replica symmetry breaking, and we show 
that the MP continuous solution ($R=\infty$, as is appropriate for 
$D<2$) does not generate negative eigenvalues. 

The general discrete expression derived here allows to control carefully 
the limiting process to the MP continuous solution. It will be also 
of use in further discussions of the solutions for the case $D>2$ 
(and in particular $D \to 4$) an unclear case to date.

In section 2 we summarize the MP approach down to the 
quadratic fluctuation terms defining the Hessian (see also 
\cite{Goldschmidt}). 
Section 3 is devoted to the introduction of a discrete Fourier-like 
transform on the replica overlaps. In section 4 we briefly discuss 
the equation of state at mean field level. In section 5 we discuss 
and exhibit the 
replicon eigenvalues. Section 6-8 are devoted to the tougher case 
of the longitudinal-anomalous eigenvalues.

\section{Summarizing MP approach}

We thus consider the MP random manifold model defined by  

   \be
     {-\cal H}(\vec{\omega})
     =
     -\frac{1}{2} 
     \int dx^D \left[
	       	     \sum_{\mu=1}^D \left(
				    	  \frac{\partial \vec{\omega}(x)}
					       {\partial x_{\mu}}
				    \right)^2 +
		     2V(x,\vec{\omega}(x)) + \mu \vec{\omega}^2(x)
	       \right] 
     \label{starting_lagrangean}
   \ee
where $d=N+D$ is the dimension of the space, $N$ the dimension of 
$\vec{\omega}$, {\it i.e.} of the tranverse space, and $D$ the dimension of 
the vector $x$, the longitudinal directions. The quenched random potential 
$V$ has a gaussian distribution of zero mean and  correlation 

    \be
      \overline{V(x,\vec{\omega}) V(x^{\prime},\vec{\omega}^{\prime})}= 
      -\delta^{(D)}(x-x^{\prime}) N 
      f\left\{
              \frac{[\vec{\omega}-\vec{\omega}^{\prime}]^2}{N}
       \right\}
    \ee
with

     \be
        f(y)=\frac{g}{2(1-\gamma)} \left( \theta + y\right)^{1-\gamma}
     \ee
where the function is regularized at small arguments by the cut-off $\theta$.

Averaging over the random quenched variable $V$ via replicas 
replaces (\ref{starting_lagrangean}) by 

	 \be
	   {-\cal H_{\rm rep}}(\vec{\omega}_a)
           =
           -\frac{1}{2} 
           \int dx^d \left[
		    \sum_{a}
		    \left\{
	       	    \sum_{\mu=1}^D \left(
				    	\frac{\partial \vec{\omega}_a (x)}
					       {\partial x_{\mu}}
			           \right)^2
			   + \mu \vec{\omega}^2_{a}(x)
	            \right\}
		    +\beta
		     \sum_{a,b} N 
		          f\left\{
                            \frac{[\vec{\omega}_a(x)-\vec{\omega}_b(x)]^2}{N}
                          \right\}
		      \right]		    
     \label{replica_lagrangean}
   \ee

In studying (\ref{replica_lagrangean}) MP have used first a variational
approach (identical to the Hartree-Fock 
approximation), which becomes exact  in the large $N$ limit. 
Then they introduced auxiliary 
fields with the advantage that it generates correction to 
Hartree-Fock, namely

\begin{itemize}
\item[$(i)$]{The zero loop term embodying the effect of quadratic 
             fluctuation, {\it i.e.} the logarithm of the determinant 
	     of the Hessian matrix}
\item[$(ii)$]{$\frac{1}{N}$ corrections in a systematic way.
              \footnote{Note that in ref.[8] the case $1-$RSB, relevant 
               for short ranged correlations, is extensively studied both for 
               its stability and its $1/N$ corrections 
                  (given in closed form).}}
\end{itemize}

In this work we are interested in $(i)$ studying the Hessian and 
its eigenvalues around the mean field (Hartree-Fock or variational) 
solutions. We leave the study of $(ii)$ for a separate 
publication\cite{Carlucci}.

Following MP we introduce the $n(n+1)/2$ component auxiliary fields

	  \be
	    r_{ab}(x)=\frac{1}{N}\vec{\omega}_a (x) \vec{\omega}_b(x) 
	  \ee
and its associated Lagrange multipliers $s_{ab}(x)$. We need to compute 
the generating function 

    \begin{eqnarray}
      \lefteqn{
      \overline{Z^n}
      =
      \int\prod_{a=1}^n d[\vec{\omega}_a] \, 
      \int\prod_{a \leq b}^n d[r_{ab}] d[s_{ab}] \times} \nonumber \\[0.5cm]
      & & \times 
      \exp\left\{
      		 -\frac{\beta}{2} \sum_{ab} \int dx s_{ab}(x) 
		 \left[N r_{ab}(x)- \vec{\omega}_a(x) \vec{\omega}_b(x) \right] 
          \right\} \times \\[0.5cm]
      & &\times 
      \exp\left\{
      		 -\frac{\beta}{2}\int dx \sum_{a=1}^n 
		 \left[
		       \sum_{\mu=1}^D \left(
				    	\frac{\partial \vec{\omega}_a (x)}
					       {\partial x_{\mu}}
			           \right)^2
			   + \mu \vec{\omega}^2_{a}(x)
	         \right] 
		 -
		 \frac{\beta^2}{2}
		 \int dx \sum_{ab} N f\left(
				     	    r_{aa}(x)+r_{bb}(x) -2 r_{ab}(x)  
				      \right)
	  \right\}    \nonumber 
     \end{eqnarray}
now quadratic in $\vec{\omega}\mbox{'s}$, hence after integrating over 
$\vec{\omega}_a$  
 
     \begin{eqnarray}
         \overline{Z^n}&=&\int \prod_{a \leq b} d[r_{ab}] d[s_{ab}] 
	 \exp\left[ N {\cal G}\{r;s\}\right] \nonumber \\[0.5cm]
	 {\cal G}\{r;s\} &=& -\frac{\beta}{2} \sum_{ab} \int dx 
	 \left[
	       s_{ab}(x) r_{ab}(x) + \beta f(r_{aa}(x)+r_{bb}(x)-2r_{ab}(x))
	 \right] + {\cal S}[s] \\[0.5cm]
	 \exp {\cal S}[s] & = & \int \prod_a d[\vec{\omega}_a] 
	 \exp\left\{
	            -\frac{\beta}{2} \sum_{ab} \int dx 
		    \left[ \vec{\omega}_a(x) \left( (-\nabla^2 +\mu)\delta_{ab} 
			   - s_{ab}(x) \right) \vec{\omega}_b(x) 
	            \right]
             \right\} \nonumber 
     \end{eqnarray}

Assuming an uniform saddle point (for large N) with 

	 \begin{eqnarray}
	    r_{ab}(x) & = & \rho_{ab} + \delta r_{ab}(x) \nonumber \\[0.5cm]
	    s_{ab}(x) & = & \sigma_{ab} + \delta s_{ab}(x) 
	 \end{eqnarray}
one can expand around it. MP thus obtain 

\begin{itemize}
\item[$(i)$]{The mean field contibution ${\cal G}^{(0)}\{r;s\}$ }
\item[$(ii)$]{The stationarity condition (vanishing of linear terms) 
	      that defines $\rho$ and $\sigma$ (identical to the 
	      variational answer\footnote{Only in the large $N$ limit. 
	      If $N$ is finite, in the variational answer $f(y)$ is 
	      replaced by $\hat{f}(y)= \frac{1}{\Gamma(N/2)} \int_0^{\infty} 
	      d \alpha \alpha^{N/2 -1} e^{-\alpha} f(2\alpha y/N)$ with 
	      $\hat{f} \to f$ as $N\to \infty$.})}
\end{itemize}

\begin{itemize}
\item[$\bullet$]{Stationarity with respect to $s_{ab}(x)$:}
\end{itemize}

	\be
	  \rho_{ab}=\frac{1}{\beta} \sum_p G_{ab}(p) 
	\ee
where 
        \be
	  \left( G^{-1} \right)_{ab}= (p^2 +\mu) \delta_{ab} - \sigma_{ab} 
	\ee

\begin{itemize}
\item[$\bullet$]{Stationarity with respect to $r_{ab}(x)$:}
\end{itemize}

	\begin{eqnarray}
	  \sigma_{ab}&=& 2\beta f^{\prime}
	               \left(\rho_{aa} + \rho_{bb} - 2 \rho_{ab}\right)
		       \\[0.5cm]
          \sigma_{aa}& =&-\sum_{b(\neq a)} \sigma_{ab}
	  \label{2_condition} 
	\end{eqnarray}

\begin{itemize}
\item[$(iii)$]{ The Hessian matrix of the quadratic form in $\delta s_{ab}$ 
		(after integration of the quadratic terms in 
		$\delta r_{ab}$ which in particular says that 
		$\delta s_{aa}=\sum_{a \neq b} s_{ab}$, {\it i.e.} } 
\end{itemize}

	\be
	  \overline{Z^n} \sim 
	  \exp\left[ N {\cal G}^0\{\rho;\sigma\}\right]
	  \int \prod_{a<b} d[ \delta s_{ab}] 
	  \exp -\frac{1}{2} \sum_{ {a<b} \atop {c<d}} \sum_p 
	  \delta s_{ab}(p) M^{ab;cd}(p) \delta s_{cd}(-p) 
	\ee

\begin{eqnarray}
	   M^{ab;cd}(p) &= &- \frac{\delta^{ab;cd}}
			    	 {2f^{\prime\prime}
				  (\rho_{aa}+\rho_{bb}-2\rho_{ab}) } - 
	                \sum_q\left\{
				     G_{ac}(q)+G_{bd}(q)-G_{ad}(q)-G_{bc}(q)\right\} \times \nonumber \\ 
&& \times				     \left\{G_{ac}(p-q)+G_{bd}(p-q)-G_{ad}(p-q)-G_{bc}(p-q)\right\} 
			 \label{Hessian}
\end{eqnarray} 

\section{Discrete Fourier Transform}

We shall work in a formalism where the level $R$ of replica symmetry breaking 
is kept discrete ($R=0$ replica symmetric, $R=1$ one step of breaking, 
$R=\infty$ Parisi breaking, {\it i.e.} continuous overlaps). In the 
end, if necessary, we shall let $R\to\infty$. 

For an observable $a_{\alpha \beta}$, with $a_{\alpha \beta}=a_r$ 
for an overlap $\alpha \cap \beta=r$, define the discrete Fourier 
transform as 

	  \be
	     \hat{a}_k=\sum_{r=k}^{R+1} p_r \left( a_r - a_{r-1} \right) 
	  \ee
and its inverse 
          
	  \be
	     a_r=\sum_{k=0}^r \frac{1}{p_k}\left(\hat{a}_k-\hat{a}_{k+1}\right)
	  \ee
Here the $p_r\mbox{'s}$ are the size of the successive Parisi boxes, 
$p_0 \equiv n$, $p_{R+1} \equiv 1$. The above transform has the 
properties 

	   \begin{equation}
	      a_r - a_{r-1} = \frac{1}{p_r} (\hat{a_r} - \hat{a}_{r+1}) 
	   \end{equation}
	  \begin{eqnarray}
	      \sum_{\gamma} a_{\alpha \gamma} b_{\gamma \beta}= c_{\alpha 
	      \beta} \\[0.5cm]
	      \hat{a}_k \hat{b}_k = \hat{c}_k 
	   \end{eqnarray} 

In particular

	      \be
	        \widehat{\left( a^{-1}\right)}_k
		=
		\frac{1}{\hat{a}_k}
	      \ee
All this is but a discrete version of the operations introduced in MP 
for the continuum ($R\to \infty$). Note that in the above all 
quantities with indices outside the range $(0,1,\dots,R+1)$  are to 
be taken identically null.
If the limit $R\to \infty$ is understood as 

	     \begin{eqnarray} 
	       \hat{a}_r &\to& \hat{a}(u) \nonumber \\[0.5cm]
	       a_r       &\to& a(u) 
	     \end{eqnarray}
then the connexion with MP is, in their notations, 

             \begin{eqnarray} 
	       \hat{a}(u)=\tilde{a}- \langle a \rangle -[a](u) \nonumber 
	       \\[0.5cm]
	       [a](u)=-\int_0^u dt a(t) + u a(u) 
	     \end{eqnarray}

\section{The equation of state}

With the above definitions we have 

     \begin{eqnarray} 
        \rho_r & = & \frac{1}{\beta} \sum_p G_r(p) \\[0.5cm] 
	\hat{G}_k(p) &=& \frac{1}{p^2 + \mu -\hat{\sigma}_k} 
     \end{eqnarray} 

and the equation of state 

    \begin{eqnarray} 
      \sigma_r = 2\beta f^{\prime}
      \left(
      	    \frac{2}{\beta} \sum_q\left[ G_{R+1}(q) -G_r(q)\right] 
      \right) & & r=0,1,\dots,R \\[0.5cm] 
      0 = - \sum_{t=0}^{R+1} p_t 
      \left( \sigma_t - \sigma_{t-1}\right) & & 
    \end{eqnarray}
The last equation, identical to  (\ref{2_condition}), also writes

    \be
      \hat{\sigma}_0=\sum_{r=0}^{R+1} p_t(\sigma_t - \sigma_{t-1})=0
    \ee
To solve for the equation of state one uses the transform that allows 
to write 

   \be
     \left[
           G_r(q) - G_{r-1}(q) 
     \right]
     =
     \frac{1}{p_r} 
     \left[
           \hat{G}_r(q) - \hat{G}_{r+1}(q) 
     \right]
   \ee
and 
    \be
      y_r \equiv \frac{2}{\beta}\sum_q
      \left[
            G_{R+1}(q) - G_{r}(q) 
     \right]=
     \frac{2}{\beta} 
     \sum_q
     \left\{
     	    \sum_{k=r+1}^R \frac{1}{p_k}(\hat{\sigma}_k - \hat{\sigma}_{k+1})
	    \hat{G}_k(q)\hat{G}_{k+1}(q) +\hat{G}_{R+1}(q) 
     \right\}
   \ee
The equation of state then writes, with 

    \[
      f^{\prime}(y_r)=\frac{g}{2} \left[ \theta + y_r \right]^{-\gamma},
    \]
 
    \be
    	\sigma_r=2\beta f^{\prime}(y_r) 
    \ee
{\it i.e.} 

    \be
      \left[
           \frac{1}{g\beta} \sum_{k=0}^r \frac{1}{p_k}
	   [\hat{\sigma}_k-\hat{\sigma}_{k+1}]
      \right]^{-\frac{1}{\gamma}}= 
      \theta + \frac{2}{\beta}
      \sum_q\left\{
      		   \sum_{r=k+1}^R \frac{1}{p_k}
                   (\hat{\sigma}_k - \hat{\sigma}_{k+1})
	           \hat{G}_k(q)\hat{G}_{k+1}(q) +\hat{G}_{R+1}(q) 
            \right\}.
   \ee
Here we have 
	
   \be
      \hat{G}_k(q)=\frac{1}{q^2+\mu -\hat{\sigma}_k}
   \ee
where $\mu$ is an infrared cut-off, which can be set to zero in appropriate 
situations. In the continuum limit  

	    \begin{eqnarray*} 
	       -\hat{\sigma}_k &\to& [\sigma](x) \\[0.5cm]
	        p_k            &\to& x 
            \end{eqnarray*}
(in the Parisi gauge) easily restituting the MP solution as given for the case 
of noise with long range correlations. 

 For later use, let us take a one step difference in the equation of state 

    \be
      \sigma_r - \sigma_{r-1}
      =
      \frac{1}{p_r}
      \left( \hat{\sigma}_r - \hat{\sigma}_{r+1} \right) 
      =
      2\beta\left[ f^{\prime}(y_r) - f^{\prime}(y_{r-1})\right]
      \label{sigma}
    \ee
with 
    \be
      y_{r-1}-y_{r}
      =
      \frac{2}{\beta} 
      \sum_q\left[G_r(q) - G_{r-1}(q)\right]
      =
 \frac{2}{\beta}   \sum_q\frac{1}{p_r}\left[\hat{G}_r(q) - \hat{G}_{r+1}(q)\right]
    \ee
{\it i.e.} 

     \be
       =
       \frac{2}{\beta}
       \left( \frac{\hat{\sigma}_r - \hat{\sigma}_{r+1}}{p_r} \right)
       \sum_q \hat{G}_r(q) \hat{G}_{r+1}(q)
     \ee
For small $y_{r-1}-y_{r}$ (infinitesimal in the continuum) one gets 

    \be
      1=-4\sum_q \hat{G}_r(q) \hat{G}_{r+1}(q)
        f^{\prime\prime}(y_r)
	\label{equation_state}
   \ee
a formula used below.

\section{Hessian Eigenvalues: the replicon sector} 

The Hessian (\ref{Hessian}) written in terms of overlaps becomes 

    \be
      M^{a b; c d}(p) 
      =
      M^{r;s}_{u;v} (p) 
      \label{four_matrix}
    \ee
with 

     \begin{eqnarray} 
        r=a \cap b &,& s=c \cap d \nonumber \\[0.2cm] 
	u = \max( a \cap c &;& a \cap d) \\[0.2cm] 
	v = \max( b \cap c &;& b \cap d) \nonumber 
     \end{eqnarray} 
In ultrametric space, only three of the four overlaps are distinct. In the 
replicon subspace $r=s$, and we have two independent overlaps 
$u,v \geq r+1$. In the longitudinal anomalous sector, if $r \neq s$, one 
can keep $\max(u,v)$ to parametrize $M$. 

For the replicon sector, and on general 
ground\cite{Kondor, deDominicis_1,Temesvari}, 
the eigenvalues are given 
by the double discrete Fourier transform 

   \be
     \lambda_p(r; k,l)
     =
     \sum_{u=k}^{R+1} p_u  
     \sum_{v=k}^{R+1} p_v
     \left( 
           M^{r;r}_{u;v} - M^{r;r}_{u-1 ;v}- M^{r;r}_{u;v-1}+ 
	   M^{r;r}_{u-1;v-1}
     \right)
     \label{R_eigenvalue}
  \ee
for $k,l \geq r+1$. 

Using the Hessian expression

   \be
     M^{r;r}_{u;v} 
     =
     -\frac{\delta^{\mbox{\tiny Kr}}_{u;R+1}\delta^{\mbox{\tiny Kr}}_{v;R+1}} 
           {2 f^{\prime\prime}_r} 
     -
     \sum_q
     	   \left[
	   	 G_u(q)+G_v(q)-2G_r(q)
           \right]
	   \left[
	   	 G_u(p-q)+G_v(p-q)-2G_r(p-q)
           \right]
   \ee

Hence, for the {\it double} transform 

   \be 
     \lambda_p(r;k,l)
     =
     -\frac{1}{2f^{\prime\prime}_r} - 
     \sum_q
          \left[
	  	\hat{G}_k(q)\hat{G}_l(p-q) + \hat{G}_l(q)\hat{G}_k(p-q)
	  \right]
   \ee
and using (\ref{equation_state}), valid in particular in the continuum,  

    \begin{eqnarray} 
      \lambda_p(r;k,l)
     &=&
     \sum_q
          \left\{
	  	2\hat{G}_r(q) \hat{G}_{r+1}(q) - 
		\left[
	  	     \hat{G}_k(q)\hat{G}_l(p-q) + \hat{G}_l(q)\hat{G}_k(p-q)
		\right]
	  \right\}  \nonumber \\[0.5cm] 
     l,k &\geq & r+1 
     \end{eqnarray}
The propagators $\hat{G}_r(q)$ are decreasing functions of $|q|$ and of 
$r$ (since $-\hat{\sigma}_r$ is an increasing function). Hence the most 
{\it dangerous} eigenvalue is 

     \be
       \lambda_p(r;r+1,r+1)
     =
     2\sum_q
          \left\{
	  	\hat{G}_r(q) \hat{G}_{r+1}(q) - 
		\hat{G}_{r+1}(q) \hat{G}_{r+1}(p-q)
	  \right\}		       
      \ee
or in the continuum limit 

   \be
     \lambda_p(x;x,x) 
     =
     2\frac{1}{(2\pi)^D} 
     \int dq^D \left\{
	       	      \frac{1}{q^2+\mu-\hat{\sigma}(x)}
		      - 
	   	      \frac{1}{(p-q)^2+\mu-\hat{\sigma}(x)}
	       \right\} \frac{1}{q^2+\mu-\hat{\sigma}(x)}
   \ee
For $D<4$, one can set the ultraviolet cut-off to infinity and rewrite 
the dangerous replicon eigenvalues, in the continuum limit, as 

    \be
      \lambda_p(x;x,x)
      =
      \sum_q\left[
                  \hat{G}_x(q)-\hat{G}_x(p-q)
            \right]^2
    \ee
where the marginal stability is obviously 
displayed\footnote{It is worth mentioning at this point that the limit 
$D\to 4$ is delicate to take. Indeed solving for the equation of state (30-32),
 one finds that the breakpoint $x_1$ 
$\left[ \dot{[\sigma]}(x)=0\,\,\mbox{for}\,\,x>x_1 \right]$ crosses its boundary value 
$x_1=1$, at some intermediate dimension $D_0(\gamma)$, $2< D_0 <4$. This 
delicate point is left out for further consideration. See also \cite{Bal} }.

\section{Hessian eigenvalues: the longitudinal anomalous sector} 

The analysis of longitudinal anomalous sector is notoriously more 
difficult. It has been considerably simplified by the 
introduction\cite{deDominicis_2,deDominicis_1,Temesvari} 
of {\it kernels}, {\it i.e.} by going to a block-diagonal 
representation of the ultrametric matrices. Take in (\ref{four_matrix})  
a component of the longitudinal anomalous sector 
$M^{r;s}_{u;v}$, $r \neq s$, which, given that only three overlaps can 
be distinct, can be written\footnote{Except for the 
components $M^{r;r}_{u;v}$ that also incorporate a replicon contribution.}

   \[
     M^{r;s}_t\,\,\,\,\,t=\max(u;v) 
   \]
To $M^{r;s}_t$ is associated a kernel (in fact a {\it Fourier} transform) 
$K^{r;s}_k$, where $k=0,1,2,\dots,R+1$, $k=0$ corresponding to the 
longitudinal sector. Likewise for the inverse $M^{-1}$ 
of the mass operator matrix we have 
a kernel $F^{r;s}_k$. The relationship between them is the Dyson's 
equation\cite{deDominicis_1, Temesvari}

	 \be
	   \hat{F}^{r;s}_k 
	   =
	   K^{r;s}_k - \frac{1}{2}\sum_{t=0}^{R+1}  
	   K^{r;t}_k\frac{\Delta_k(t)}{\Lambda_k(t)}
	   \hat{F}^{t;s}_k 
	   \label{LA_dyson}
	 \ee

	 \be
	   \hat{F}^{r;s}_k 
	   =
	   -\Lambda_k(r)
	    F^{r;s}_k 
	    \Lambda_k(s)
	   \label{LA_dyson_2} 
	 \ee
Here the $\Lambda_k(r)$ are special replicon eigenvalues defined by 

     \be
       \Lambda_k(r)=
       \left\{ \begin{array}{cc}
	        \lambda(r;k,r+1) & k \geq r+1 \\[0.5cm] 
		\lambda(r;r+1,r+1) & k < r+1
		\end{array} 
       \right. 
       \label{special_R_eigenvalue}
     \ee
with $\lambda$ as in (\ref{R_eigenvalue}) and

     \be
       \Delta_k(t)=
       \left\{
       	      \begin{array}{cc}
	       \frac{1}{2}(p_t - p_{t+1}) & t<k-1 \\[0.5cm] 
               \frac{1}{2}(p_t - 2 p_{t+1}) & t= k-1 \\[0.5cm] 
               (p_t - p_{t+1}) & t>k-1 
	      \end{array}
        \right.
      \ee
The kernel $K^{r;s}_k$ itself is a transformed of the mass operator matrix 
$M^{r;s}_t$. It writes as a generalized discrete Fourier transform 

      \be
        K^{r;s}_k
	=
	\left( 
	       \sum_{t=k}^r + 2 \sum_{t=r+1}^s + 4 \sum_{t=s+1}^{R+1}
	\right) p_t \left(M^{r;s}_t - M^{r;s}_{t-1}\right)
      \ee
where we have displayed the case $k \leq r \leq s$ (for different 
orderings the summation $\sum_{t=k}$ is only kept in the allowed summands). 
Knowing $K^{r;s}_k$, one then computes $F^{r;s}_k$ {\it via}  
(\ref{LA_dyson}, \ref{LA_dyson_2}) and $(M^{-1})^{r;s}_t$ {\it via} an 
inverse transform. That is if one is able to solve the integral 
equation (\ref{LA_dyson}) which is only possible with especially 
simple kernels $K^{r;s}_k$. This turned out to be the case when 
studying the bare propagator of the standard spin 
glass\cite{deDominicis_1} where 

	 \be
\frac{1}{4}	   K^{r;s}_k=A_k(min(r;s))
	   \label{sg_kernel}
	 \ee
The same functional form also appears for the spin glass model 
introduced by Th.Nieuwenhuizen\cite{Nieu}. 
For the system studied here, let us construct the kernel $K^{r;s}_k$ 
with the matrix $M^{r;s}_t$ as given by the Hessian (\ref{Hessian}) 
manifestly symmetric in $r,s$. Let us choose $r<s$ and we then find 

   \begin{eqnarray} 
 \frac{1}{4}    K^{r;s}_k(p) &=& 
     -\sum_q 
	             \sum_{t=\max(s+1,k)}^{R+1} p_t
	\left\{	     \left[
		     	   G_t(q) G_t(p-q)-G_{t-1}(q) G_{t-1}(p-q)
	             \right]-
	      \right. \nonumber \\[0.6cm] 
    &&\mbox{}- 
          G_s(q)
	  \left[
		 	  G_t(p-q)- G_{t-1}(p-q) 
		    \right]-
     \left.
	   \left[
		    G_t(q)- G_{t-1}(q) 
	      \right]G_s(p-q)
       \right\}
     \label{LA_kernel}
    \end{eqnarray}
which also rewrite as 

      \be
\frac{1}{4}        K^{r;s}_k
	=
	\left\{
	       \begin{array}{cc}
	        B_k(s) & r< s< k-1 \\[0.5cm] 
		B_{s+1}(s) & s \geq k-1 
	       \end{array}
        \right.
	\label{B_kernel}
      \ee
where

	\be
	  B_k(s) 
	  =
	  -\sum_q
	  \left\{
	  	 \left(\widehat{G(q)G(p-q)}\right)_k 
		 -
		 G_s(q) \hat{G}_k(p-q)
		 -
		 \hat{G}_k(q)G_s(p-q)
          \right\}
        \ee
The important point is that the kernel only depends on $max(r,s)$ 

    \be
\frac{1}{4}       K^{r;s}_k = B_k(\max(r;s))
    \ee
thus rendering soluble the integral equation (\ref{LA_dyson}).

\section{Solution of the Dyson equation: the propagator kernel}

In the case of $\frac{1}{4}K^{r;s}_k=A_k(\min(r;s))$ the solution of the 
Dyson equation (\ref{LA_dyson}) has been given (in the continuum 
limit) in \cite{deDominicis_1}. We wish to extend it here to the case 
$\frac{1}{4}K^{r;s}_k = B_k(\max(r;s))$. In this section we work for a given value 
of $k$, {\it i.e.} we remain in a sub-block of the longitudinal 
anomalous sector ($k=0$ for the longitudinal).

The propagator kernel of eqs (\ref{LA_dyson}, \ref{LA_dyson_2}) is 
obtained as 

	 \be
	   \frac{1}{4} F^{r;s}_k
	   =
	   \frac{\phi^+_k(\min (r;s)) \phi^-_k(\max (r;s))}
	        {\Lambda_k(r) \Delta_k \Lambda_k(s)}
         \ee
where $\Delta_k$ is a {\it Wronskian} and $\phi^{\pm}_k$ the regular 
and irregular solution of a discrete Sturm-Liouville equation and 
$\Lambda_k$ the special replicon eigenvalues of 
(\ref{special_R_eigenvalue}). 
Namely we have for the Wronskian 

       \be
         \Delta_k
	 =
	 \frac{\phi^-_k(s) {\cal D}_s \phi^+_k(s) - 
	       \phi^+_k(s) {\cal D}_s \phi^-_k(s)}
	      { {\cal D}_s B_k(s)}
       \ee
with 
       \be
           {\cal D}_s f(s) \equiv f_{s+1}-f_s 
       \ee

The functions $\phi^{\pm}$ are given by 
     
    \be
      \frac{1}{2}\phi_k^+(s)
      = 
      1 + \sum_{t=0}^s\left[B_k(t)-B_k(s)\right]
      \frac{\Delta_k(t)}{\Lambda_k(t)} \phi_k^+(t)
      \label{phi_plus} 
    \ee

    \be
      \frac{1}{2}\phi_k^-(s)
      = 
      - \sum_{t=s}^R\left[B_k(t)-B_k(s)\right]
      \frac{\Delta_k(t)}{\Lambda_k(t)} \phi_k^-(t)
      - B_k(s)C_k 
    \ee
with $B_k(s)$ as in (\ref{B_kernel}). Note that, contrary to the solution 
obtained for (\ref{sg_kernel}), where $C_k=1$, here, $\phi^-_k$ and 
$\phi^+_k$ remain coupled {\it via} $C_k$ 

	   \[
	     C_k 
	     \equiv 
             \sum_{t=0}^R
	     \frac{\Delta_k(t)}{\Lambda_k(t)} \phi_k^+(t) 
	   \] 
In terms of $\phi^+_k$ the Wronskian can also be written as 

           \be
	     \Delta_k
	     =
	     4 C_k
	     \left[
	           1 + \sum_{s=0}^R B_k(s)
	           \frac{\Delta_k(s)}{\Lambda_k(s)} \phi_k^+(s) 
	     \right].
	   \ee

\section{Stability discussion in the LA sector} 

Given the explicit expressions (\ref{B_kernel}) of 
$B_k(s)$ in terms of $\hat{G}_t$ (or $G_t$) itself a positive 
definite function, decreasing with increasing $t$ 

	\be
	  {\cal D}_t G_t
	  =
	  \frac{1}{p_{t+1}} 
	  \frac{ \hat{\sigma}_{t+1}-\hat{\sigma}_{t+2}     } 
	       { \hat{G}_{t+1}^{-1} \hat{G}_{t+2}^{-1} }  
        \ee

one can easily show that 

    \begin{eqnarray} 
     (i)  & & B(r) <0 \label{B_condition} \\[0.5cm] 
     (ii) & & {\cal D}_r B(r)\equiv B(r+1)-B(r)>0 
    \end{eqnarray}
Under these conditions we can now give a {\it lower bound} 
for the longitudinal anomalous eigenvalues. We can write out the 
determinant of the eigenvalues restricted to the sub-block $k$ : 


   \be
      \det\left( M^{(k)}_{rs} - \lambda I \right) 
      =
      \prod_{r=0}^R \left[ \Lambda_k(r) -\lambda \right]
      \left[
            1+\sum_{t=0}^R B_k(t)\frac{\Delta_k(t)}{\Lambda_k(t) -\lambda} 
	    \phi_k^+(t)
      \right]
      \label{determinant}
   \ee
which is directly related to the 
Wronskian (apart from the $C_k$ factor).  

Note that, from equation (\ref{phi_plus}), one also obtains

     \be
       {\cal D}_s \phi_k^+(s)
       =
       -2 {\cal D}_s B_k(s)
       \sum_{t=0}^s  \frac{\Delta_k(t)}{\Lambda_k(t) -\lambda} 
       \phi_k^+(t).
     \ee
With the boundary value $\phi_k^+(0)=2$ and ${\cal D}_s B(s)>0$, one 
infers that, in case of  
 $\lambda < \Lambda_k(t)$ for all $t$,

       \begin{eqnarray}
         \phi^+_k(s) & > & 0 \label{phi_condition} \\[0.2cm]
	 {\cal D}_s \phi^+_k(s) & > & 0 
       \end{eqnarray} 
Consider now in equation (\ref{determinant}) a small value of $\lambda$ 
such that
 $\lambda < \Lambda_k(t)$ for all $t$
. For that value $\lambda$ 
the determinant of (\ref{determinant}) {\it cannot vanish}, it 
is always positive  on account of  (\ref{B_condition}, \ref{phi_condition}). 
Hence the spectrum of the longitudinal anomalous sector is 
certainly bounded by the bottom values of the replicon spectrum 
$\lambda_{p=0}(r;r+1,r+1)$. This establishes stability, a marginal 
stability as in the continuum limit where $\lambda_{p=0}(x;x,x)=0$.

\section{Conclusions}

We thus have set up for random manifolds the formalism to discuss 
the stability of mean field solutions. For the well established  
continuous ($R=\infty$) solution, adequate for $D<2$, we are able to prove its 
marginal stability. The formalism used here allows a discussion of any type 
(discrete, continuous or mixed) of  mean field solutions.

It is worth noticing that, when discussing the stability, the only 
place where properties of the noise correlation function may play  
a direct role (outside of the fully continuous $R=\infty$ solutions) 
is in (\ref{sigma}-\ref{equation_state}). There the equation of state 
is used to find the alternative analytical form for $f^{\prime\prime}$, 
that cast in evidence the non-negative character of the replicon 
eigenvalues.

\section*{Acknowledgements} 
One of us (CD) gladly acknowledges useful discussions with M.Mezard.

\end{document}